\documentclass[reprint,superscriptaddress,amsmath,amssymb,aps]{revtex4-2}

\usepackage{graphicx}% Include figure files
\usepackage{dcolumn}% Align table columns on decimal point
\usepackage{bm}% bold math

\usepackage{amssymb}

\usepackage{tabularx}
\usepackage{graphicx}% Include figure files
\usepackage{dcolumn}% Align table columns on decimal point
\usepackage{bm}% bold math
\usepackage{amsmath}
\usepackage{xcolor} % <======================================= for green
\usepackage[pagebackref]{hyperref} % <====================== last packge to be called
\hypersetup{
	colorlinks   = true, %Colours links instead of ugly boxes
	urlcolor     = blue, %Colour for external hyperlinks
	linkcolor    = blue, %Colour of internal links
	citecolor   = blue %Colour of citations, could be ``red''
}
\usepackage{graphicx}
\usepackage{physics}
\usepackage{mathtools}
\usepackage{dsfont}
\usepackage{natbib} %bilbiography
\begin{document}

\preprint{APS/123-QED}
\title{Emulating a quantum Maxwell's demon with non-separable structured light}% Force line breaks with \\
\author{Edgar Medina-Segura}
\affiliation{Centro de Investigaciones en Óptica, A.C., Loma del bosque 115, Lomas del Campestre, 37150, León, Gto., México}
\author{Paola C. Obando }
%\email{paola.conchaobando@wits.ac.za}
\affiliation{School of Physics, University of the Witwatersrand, Private Bag 3, Wits 2050, South Africa}
\author{Light Mkhumbuza}
\affiliation{School of Physics, University of the Witwatersrand, Private Bag 3, Wits 2050, South Africa}
%\author{Janice A. Nair }
%\affiliation{School of Physics, University of the Witwatersrand, Private Bag 3, Wits 2050, South Africa}
\author{Enrique J. Galvez}
\affiliation{Department of Physics and Astronomy, Colgate University, Hamilton, New York 13346, USA}
\author{Carmelo Rosales-Guzmán}
\affiliation{Centro de Investigaciones en Óptica, A.C., Loma del bosque 115, Lomas del Campestre, 37150, León, Gto., México}
\author{Gianluca Ruffato}
\affiliation{Department of Physics and Astronomy, University of Padova, Via Marzolo 8, 35131, Padua, Italy}
\author{Filippo Romanato}
\affiliation{Department of Physics and Astronomy, University of Padova, Via Marzolo 8, 35131, Padua, Italy}
\author{Andrew Forbes}
\affiliation{School of Physics, University of the Witwatersrand, Private Bag 3, Wits 2050, South Africa}
\author{Isaac Nape}
\email{isaac.nape@wits.ac.za}
\affiliation{School of Physics, University of the Witwatersrand, Private Bag 3, Wits 2050, South Africa}

\date{\today}% It is always \today, today,

\begin{abstract}

  \noindent Maxwell's demon (MD) has proven an instructive vehicle by which to explore the relationship between information theory and thermodynamics, fueling the possibility of information driven machines. A long standing debate has been the concern of entropy violation, now resolved by the introduction of a quantum MD, but this theoretical suggestion has proven experimentally challenging. Here, we use classical vectorially structured light that is non-separable in spin and orbital angular momentum to emulate a quantum MD experiment.  Our classically entangled light fields have all the salient properties necessary of their quantum counterparts but without the experimental complexity of controlling quantum entangled states. We use our experiment to show that the demon's entropy increases during the process while the system's entropy decreases, so that the total entropy is conserved through an exchange of information, confirming the theoretical prediction. We show that our MD is able to extract useful work from the system in the form of orbital angular momentum, opening a path to information driven optical spanners for the mechanical rotation of objects with light.  Our synthetic dimensions of angular momentum can easily be extrapolated to other degrees of freedom, for scalable and robust implementations of MDs at both the classical and quantum realms, enlightening the role of a structured light MD and its capability to control and measure information. 

\end{abstract}

\maketitle
%\tableofcontents
\section{Introduction}

\noindent Maxwell’s demon (MD), a thought experiment proposed by James Clerk Maxwell, remains as a highly topical subject in both theoretical and experimental physics \cite{rex2017maxwell}.  In its original casting, the demon is able to exploit useful information of a thermodynamic system and trades this for thermodynamic advantage, e.g., to perform useful work.  Maxwell's demon lies at the heart of the interrelation between quantum
information and thermodynamics, a topic of theoretical interest \cite{maxwell2001,Zurek2003,Kim2011,leff2002maxwell,colloquium09}, resulting in notable progress in our understanding of the relation between thermodynamics and information \cite{bennett1982,parrondo2015}. In parallel with this theoretical progress, technological advances have contributed to new experimental techniques for manipulating small systems, thereby allowing experimental demonstrations of Maxwell’s demon using superconducting qubits \cite{cottet2017}, microscopic particles with feedback control \cite{toyabe2010}, colloidal particles \cite{berut2012}, nuclear magnetic resonance(NMR) spectroscopy  \cite{camati2016} as well as electronic MDs with single-electrons \cite{koski2014exp2,koski2015}.  Among its notable applications are feedback control at both classical and quantum level \cite{sagawa2008}, power generation \cite{chida2017}, nanoscale information-processing devices \cite{hanggi2009}, superconducting circuit quantum electrodynamics (QED)  demons \cite{masuyama2018}, photonic fiber based systems \cite{zanin2022enhanced} and information-powered engines, for instance, converting information to free energy \cite{toyabe2010}.

The introduction of a quantum MD allows for the entire thermodynamical cycle and information gain and loss to be treated within a unified quantum framework \cite{lloyd97}, with quantum implementations demonstrated with single electrons \cite{koski2014} and single photons \cite{vidrighin2016}.  Although entanglement of the system and the demon is crucial in the information extraction and erasure process, the only qubit example of a quantum MD is that of local entanglement of superconducting qubits \cite{cottet2017}, requiring a complex experimental set-up.  We propose the use of vectorially structured light to implement an optical version of a quantum Maxwell demon. Structured light arises from the ability to tailor the degrees of freedom of light \cite{forbes2021structured,piccardo2021roadmap,he2022,Nape2023Q_highdim}. Importantly, it is possible to construct classical light fields that are non-separable in two or more degrees of freedom with all the non-separability traits of quantum (non-separable) entangled states but without non-locality, sometimes referred to as classical entanglement \cite{spreeuw1998classical,Konrad2019,forbes2019classically,shen2022nonseparable,eberly2016quantum}. Such non-separable structured light provides a platform to perform operations on two synthetic dimensions of classical light as if they were qubits in the quantum regime, and has proven powerful in blurring the classical-quantum divide \cite{ndagano2017characterizing,guzman2016demonstration,di2023ultra,nape2022revealing,qian2015shifting,wang2024structured}. 

Here, we exploit the aforementioned approach using the spin angular momentum (SAM) and orbital angular momentum (OAM) states of light as our non-separable degrees of freedom, allowing us to emulate a quantum version of Maxwell's demon using non-quantum light. We use our experiment to show that the quantum demon's entropy increases during the  process, from $S_D \approx 0.0731$ initially, to $S_D \approx 0.999$ and back to $S_D \approx 0$ . As as a consequence, the system's OAM states are initially in a superpositon state, to only one OAM state remaining in the final stage. This confirms that our MD is able to extract useful work (or equivalently information) from the system in the form of OAM.  Our approach has all the hallmarks of the proposed experiments for a quantum MD but at far lower experimental complexity, and marks the first optical implementation with classically entangled light. 

\section{Maxwell's Demon Concept}
\label{subsec:Maxwell demon concept}

\begin{figure*}[t]
\centering
\includegraphics[width=0.98\linewidth]{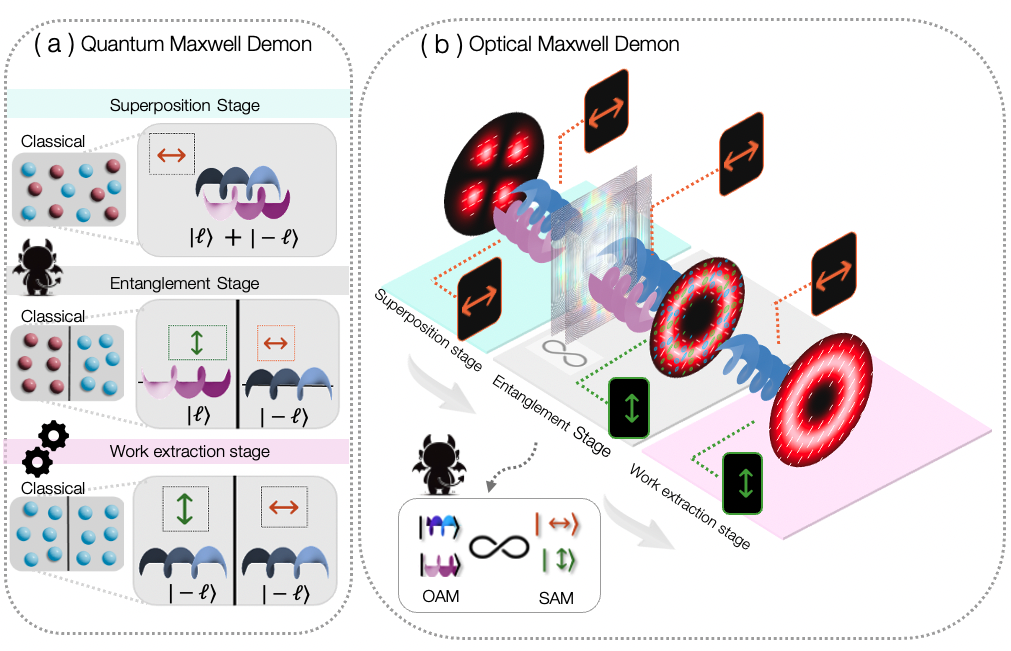}
\caption{\textbf{A quantum Maxwell's demon with classically entangled light.} \textbf{(a)} Representation of the Maxwell’s demon thought experiment, showing the demon's protocol for work extraction and control's heat transfer by using information about the microscopic state of gas molecules \cite{leff2002maxwell}. The process is divided into three stages: Superposition stage, Entanglement stage and Work extraction stage. The multi-particle system and demon in the classical scenario (depicted as a small block at the back of each stage) can be viewed as a two-level system on a quantum regime \cite{lloyd97}. This analogy can be simulated with non-quantum light, using a two level OAM system  $\{\ket{-\ell}, \ket{\ell} \}$ and SAM demon $\{H, V \}$ . Note that, the physical separation in the classical MD is now represented as two polarisation states, which are also interpreted as the demon's memory. \textbf{(b)} Conceptual simulation of Maxwell’s demon in the angular momentum scenario, using structure light. In the Superposition stage, a superposition state $\ket{-\ell} + \ket{\ell}$ is encoded in the initial beam, with the horizontal polarization. In the subsequent Entanglement stage, using a set of optical devices, a non-separable vector beam is created between OAM and SAM states. Finally, the demon flips one of the OAM states, leading to a determined OAM state and a superposition polarization states.}
\label{fig:conceptual}
	\end{figure*}

\noindent To benefit the reader, we briefly review the concepts of the traditional Maxwell's demon thought experiment. Imagine an initial state where a box contains a gas in thermal equilibrium. Now, suppose a wall with a tiny door is inserted to separate the box in two partitions. Maxwell's demon is envisioned as a microscopic, intelligent being who operates the door to filter particles based on their energy. This results in cold gas molecules accumulating in one section and hot gas molecules in the other. As a result, without any investment of work, the demon increases the temperature of one section while decreasing the temperature of the other, seemingly violating the second law of thermodynamics. This apparent reduction in entropy, achieved without any energy input, suggests a potential contradiction to the second law, which states that entropy in an isolated system can never decrease \cite{maxwell2001,leff2002maxwell}. Furthermore, since Maxwell's demon drives the system out of equilibrium, it can exploit the engineered temperature difference to perform useful work, such as cooling the hot section of the system (See the entire process illustrated in the small blocks of Fig.~\ref{fig:conceptual}(a), which are labeled as ``classical"). This begs the question: is there a violation of entropy in this process?.

A particularly relevant approach to addressing this question is found at the quantum level. A quantum MD is able to demonstrate that such violations never occur \cite{lloyd97}.  In this scenario, one can again separate the steps into three parts: first at the  Superposition stage, an initial state is prepared where the system is in a superposition state of some degree of freedom, indicating a high degree of uncertainty, while the demon's initial state is well defined in another degree of freedom. In the Entanglement stage, the demon acquires information about the system and, by applying a conditional operation, is able to sort out the system's states by becoming entangled with the system. One can imagine the demon, equipped with a two state memory, recording each of the system's states in a separate memory partition (``boxes" made of orthogonal states of some degree of freedom). At the end of the entanglement process, the interaction between system and demon produces a maximally entangled state, they become entangled, a distinctive trait of quantum demons. As a result of establishing perfect quantum correlations between the two subsystems, each is mapped onto a maximally mixed state introducing decoherence locally. Though this increases the quantum entropy of the demon's memory (and the system), the perfect correlations enable the demon to gain information about the system's states, establishing the correspondence between thermodynamic quantities and quantum information. Therefore the local Von Neumann entropy (of each subsystem) and the mutual information between the the two subsystems are equivalent. Finally, with access to the system's states, the demon can perform work on one of them. As a result, in the Work extraction stage the system reaches a definite state, while the demon's memory is left in a superposition state. 

Mathematically, this can be seen as a sequence of operations described as follows: (i) starting with an initial separable state, (ii) a conditional mode-flip operation is applied, resulting in an entangled state between the modes, and (iii) finally extract or generate work from the system, another conditional mode-flip operation is performed, leading to a disentangled state between the modes. Readers familiar with quantum computation will recognize these conditional flip operations as analogous to the quantum  ‘controlled NOT’ (CNOT) operation, which transitions between entangled and separable states depending on the input.

%\AF{xxx mathematically this can be seen as a sequence of operations defined by ... xxx An interesting fact about the spin-flip operation is that this conditional operation is known in quantum computation as the ‘controlled NOT’ (CNOT) operation }

A brief examination of the overall process shows that an initial superposition state can be transformed into a definite state through the action of Maxwell's demon by using entanglement as a resource to gain quantum information required to perform such a mapping. The demon must pay a price, reflected in the increase of its own entropy, which is then later reduced by the work extraction step. Consequently, at the end of these processes, the demon's memory —initially in a definite state— ends up in a superposition state. Therefore, no violation of the second law occurs, as the changes in the system's entropy only occur locally, keeping the total entropy of the combined system (system and demon) conserved throughout the process.

Motivated by this framework, in which a multi-particle system of the original paradox is replaced by a quantum two-level system controlled by a quantum demon (also represented by a two-level system), we propose a quantum MD using structured light. Specifically, we use the angular momentum  of light to construct the two-level states of both the system and the demon.
%\Edgar{as seen in right boxes of Fig.~\ref{fig:conceptual}(a)  where three-dimensional spirals represent helical phases associated with optical fields that carry OAM and the double arrows are polarisation states. For the experimental implementation we decided to use the linear base of polarisation but we could use the circular polarisation base which is associated with spin angular momentum (SAM), that is why we refer to the demon degree of freedom as SAM and we termed the whole experiment as Maxwell's demon with optical angular momentum.}

\section{Maxwell's demon with optical angular momentum}
\noindent Now we extend the notion to an optical Maxwell demon using the angular momentum of light, first conceptually and then by experimental implementation. 

\subsection{Optical Maxwell's Demon Concept}
A structured light beam with an azimuthal phase dependence of the form $\exp(-i\ell\phi)$, carries an OAM of $\ell \hbar$, where $\ell$ is the azimuthal quantum number (or topological charge) and $\hbar$ is Planck's constant. Since OAM states of different $\ell$ are orthogonal, it is possible to construct a basis where each beam's OAM subspace is spanned by $\mathcal{H}_2=\{ \ket{\ell}, \ket{-\ell}\}$. The detected photon intensity distribution in classical light is symmetrical and uniform in the azimuth for both basis states, with each state possessing azimuthal helicity of the same magnitude but opposite sign. These characteristics allow OAM modes with opposite helicities to be treated as a two-level system, indistinguishable in terms of intensity.  Likewise, a SAM basis can be constructed, where each beam's polarisation subspace is spanned by $\mathcal{H}_2=\{ \ket{H}, \ket{V}\}$ (we will frame our spin states in terms of linear polarisation states in anticipation of how the experiment was performed). The orthogonality of the SAM states allows the construction of a two level state. Both OAM and SAM corresponds to a two-level basis, allowing a quantum MD to be reinterpreted using this angular momentum framework, see Fig.~\ref{fig:conceptual}(a). Following the three stages of a quantum MD described in the previous section, the conceptual equivalence between the quantum MD and the optical Maxwell's demon at each stage is illustrated in Fig.~\ref{fig:conceptual}(b) and corresponds to the detailed process outlined below.

\textbf{Superposition Stage.} The initial system state is prepared as a structured light field carrying a superposition of OAM modes, given by $\ket{\phi_i}_{s}=\frac{\ket{\ell}_{s}+e^{i\theta}\ket{-\ell}_{s}}{\sqrt{2}}$, while the initial demon's memory state $\ket{\psi_i}_{d}$ is prepared as a well defined horizontal polarisation state. Thus, the initial state of the full system $\ket{\Psi_i}_{sd}$ is described by:
\begin{align}\label{eqn:in}
\nonumber\ket{\Psi_i}_{sd}&=\ket{\phi_i}_{s}\otimes \ket{\psi_i}_{d}\\
&=\frac{\ket{\ell}_{s}\ket{H}_{d}+e^{i\theta}\ket{-\ell}_{s}\ket{H}_{d}}{\sqrt{2}},
\end{align}
where $\theta$ is a general intermodal phase between the OAM states. By analogy with a quantum state, the initial state can be viewed as a product state between the system and the demon’s memory, corresponding to the OAM and SAM subspaces, respectively. The system is in a superposition while the demon’s memory is in a determined state. At this stage, the demon does not  have  knowledge about system's states since the total system is a separable state making them independent.\\

\textbf{Entanglement Stage.}  At this stage, the demon acquires information about the system. However, instead of directly \textit{measuring} the OAM states, the demon sorts the OAM states using the polarisation degree of freedom. To achieve this, the demon applies the following conditional operation: if the OAM is in the $\ket{-\ell}$ state, it performs a spin-flip operation changing its initial polarisation state from  $\ket{H}$ to $\ket{V}$. Otherwise, if the OAM is in the $\ket{\ell}$ state, the initial polarisation state remains unchanged. This conditional operation is performed by a polarisation beam splitter (PBS) as described below in section \ref{sec:exp}. The intermodal phase $\theta$ is set to $\theta=0$ during the process. Thus, the state resulting from the conditional operation is the outcome of the mapping transformation of Eq. (\ref{eqn:in}) onto the entangled state, described as follows

 \begin{equation}\label{eq:non-sep}
    \frac{\ket{\ell}_{s}\ket{H}_{d}+\ket{-\ell}_{s}\ket{H}_{d}}{\sqrt{2}} \;\; \longrightarrow \;\;\frac{\ket{\ell}_{s}\ket{H}_{d}+\ket{-\ell}_{s}\ket{V}_{d}}{\sqrt{2}}. 
\end{equation}

%\begin{equation}\label{eq:non-sep}
%\ket{\psi}_{sd_2}=\frac{\ket{\ell}_{s}\ket{H}_{d}+\ket{-\ell}_{s}\ket{V}_{d}}{\sqrt{2}}. 
%\end{equation}

In the above  transformation,  we observe that at the end of the entanglement process, a non-separable state is obtained: our classically entangled vector beam that is non-separate in two degrees of freedom (local entanglement). The system and demon are now entangled.  Similar to a quantum demon that leverages entanglement to sort out system states, our angular momentum demon achieves the sorting process by utilizing the spin-orbit coupling  of non-quantum light, recording the system's information in its two memory states. Through this process, the quantum information acquired by the demon increases since the demon's memory is perfectly correlated with the systems states, though each isolated system is a completely mixed state. \\

%\Edgar{[It is convenient change to  spin-orbit non-separability?]}}

%Note that we have set aside the label $\ket{psi}_{sd_1}$ for the full system state because there is an intermediate state, explained in the next subsection, obtained before getting the $\ket{\psi}_{sd_2}$.\\

\textbf{Work Extraction Stage.}  Finally, the demon is ready to perform useful work from the system. The demon uses the information acquired to sort OAM based on polarisation and performs the following conditional operation:  If the demon's memory is in the 
$\ket{V}$ state, a flipping operation is applied, transforming the system state from 
$\ket{-\ell}$ to $\ket{\ell}$. Conversely, if the demon's memory is in the 
$\ket{H}$ state, the system's state remains unchanged. This conditional operation can be seen as a mapping that acts on the output entangled state of Eq.~(\ref{eq:non-sep}), resulting in a final state describe by the following transformation 
\begin{equation}\label{eq:final}
   \frac{\ket{\ell}_s \ket {H}_d+\ket{-\ell}_{s}\ket{V}_d}{\sqrt{2}} \;\; \longrightarrow  \frac{\ket{\ell}_s\otimes (\ket{H}+\ket{V})_d}{\sqrt{2}}.
\end{equation}
The resulting state of this mapping is the product state (disentangled state)  $\ket{\Psi_f}_{sd}=\ket{\phi_f}_{s}\otimes \ket{\psi_f}_{d}$, where the system is in a definite state  $\ket{\phi_f}_{s}=\ket{\ell}_s$, and the demon’s memory is in a superposition state $\ket{\psi_f}_{d}=\frac{1}{\sqrt{2}}(\ket{H} +\ket{V} )_d$. Work extraction can be identified with the flipping process of the OAM state. To extract energy from the system, the demon decreases the quantum entropy (the quantum information of the subsystems) by disentangling the two subsystems, thereby paying the cost of the energy extraction by leaving the demon in a superposition state.

%%%%%%%%%%%%%%%%%%%%%%%%%%Simulations and experiement%%%%%%%%%%%%%%%%%%%%%%%%
%%%%%%%%%%%%%%%%%%%%%%%%%%%%%%%%%%%%%%%%%%%%%%%%%%%%%%%%%%%%%%%%%%%%%%%%%%%%%

%%%%%%%%%%%%%%%%%%%%%%%%%%%Figure3%%%%%%%%%%%%
%%%%%%%%%%%%%%%%%%%%%%%%%%%%%%%%%%%%%%%%%%%%%%
%From top to bottom, Experimental results. Left:(center)(right) . 

\subsection{Experimental Set-up}\label{sec:exp}

%{The experimental setup is shown schematically in Fig.~\ref{fig:opticalexp}. OAM Measurement was performed using a mode sorter composed of an unwrapper (UW), a phase corrector (PC), and a Fourier transforming lens (FTL) \cite{filippo2018}, as shown in the purple box labeled `` OAM Measurement'' in Fig.~\ref{fig:opticalexp}. Meanwhile, polarisation measurement utilized Stokes polarimetry, which was spatially resolved through a polarisation camera and the QWP$_2$, as depicted in the ``Polarisation Measurement" box in the same figure. Note that all points of interest in the experiment are marked with red dots along the laser path, indicating the output of each stage outlined here.
%%%%%%%%%%%%%%%%%%%%%%%%%%%%%%%%%%%%%%%%%%%%%%%%%
%%%%%%%%Figure2%%%%%%%%%%%%%%%%%%%%%%%%%%%%%%%%%%
\begin{figure*}[t]
\centering
\includegraphics[width=1\linewidth]{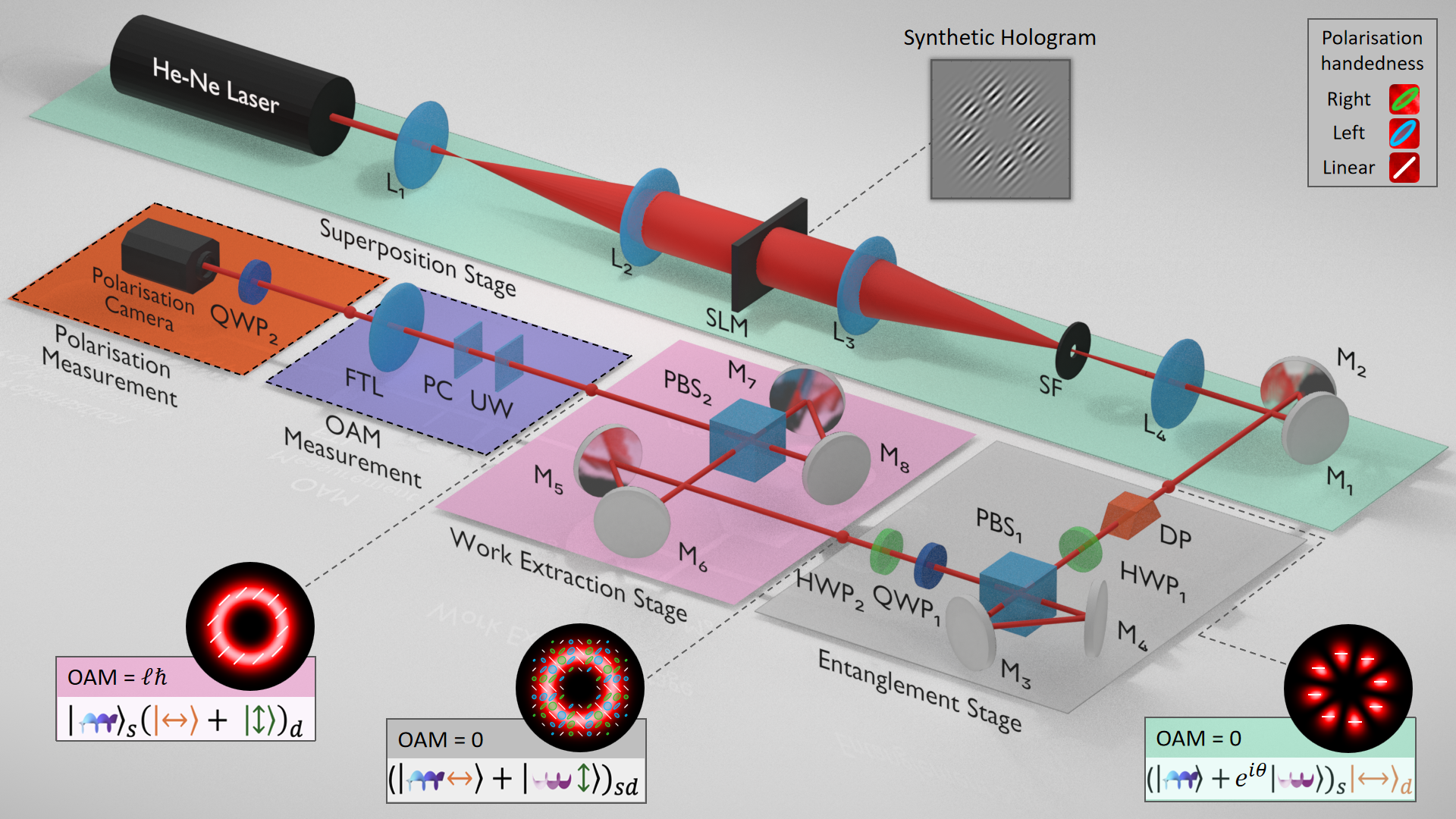}
\caption{\textbf{Experimental setup of the angular momentum  Maxwell's demon.} The experiment is divided into three stages: Superposition Stage, Entanglement Stage and Work Extraction Stage. Superposition Stage starts with the collimation of a He-Ne laser and concludes with the creation of the superposition state described by Eq. (\ref{eqn:in}). Next, in the Entanglement Stage, the PBS$_1$ along with the mirrors $M_3$ and $M_4$ are used to create a non-separable state of light in the spatial and polarisation degrees of freedom. At last, in the Work Extraction Stage, unitary transformations are applied to flipping one of the OAM states, thereby getting a state able to perform useful work. $L_i=$ Lenses, SLM = Spatial Light Modulator, SF$=$ Spatial Filter, $M_i=$ Mirrors, DV$=$Dove Prism, HWP$_i=$ Half-Wave Plates, QWP$_i=$ Quarter-Wave Plates, PBS$_i=$ Polarising Beam Splitters, UW = Unwrapper, PC = Phase corrector, FTL = Fourier transforming lens.}
\label{fig:opticalexp}
	\end{figure*}

 To experimentally validate the optical version of the quantum MD we followed the experimental schematic of Fig ~\ref{fig:opticalexp}. The experiment consists of the generation stage, \textbf{Superposition Stage}, where a scalar Gaussian beam from Helium Neon (HeNe) laser (wavelength $\lambda = 633$ nm) was expanded and collimated  using a two-lens magnifying system $L_1$ and $L_2$  to illuminate a reflective HOLOEYE PLUTO-2.1  spatial light modulator (SLM). The scalar Gaussian mode was then modulated using a Laguerre-Gaussian (LG) complex-amplitude hologram with superposition of OAM of $\ell=[-4,4]$, with a general intermodal phase of $\theta$ between the modes, encoded on the SLM, thus producing an LG mode with petal-like structure as shown in the superposition stage of Fig~ \ref{fig:opticalexp}.  Since the SLM produces multiple diffraction orders, a spatial filter or aperture (SF) was placed in the focal plane of $L_3$ to select only the positive first order and filter out the remaining orders. The selected first order mode was then imaged onto the \textbf{Entanglement stage} shown by the grey block where the beam is first passed through a Dove prism to adjust the intermodal phase to $\theta=\pi/2$ which is the mandatory value for the next steps to work properly. We then employ a half-wave plate (HWP$_1$) with its fast axis at $(-45^o/2)$ to rotate the polarisation state of the mode from linear horizontal to linear diagonal. Next, the polarising beam splitter (PBS$_1$) reflects the vertical component of the beam and flips the sign of OAM while the horizontal component is transmitted to the mirrors M$_3$ and M$_4$ preserving the original OAM state. At the output of PBS$_1$, both modes are superimposed, resulting in the intermediate state $\ket{\Psi_{int}}_{sd}$, as follows
\begin{align}
    \nonumber \ket{\Psi_{int}}_{sd}&=\frac{(\ket{H}+i\ket{V})\ket{\ell}+i(\ket{H}-i\ket{V})\ket{-\ell}}{2},\\
    &=\frac{\ket{L}\ket{\ell}+i\ket{R}\ket{-\ell}}{\sqrt{2}},
\end{align}
where $\ket{R}$ and $\ket{L}$ are the right- and left-circular polarisation states. This state is then passed through a QWP$_1$ at $45^o$ to create the vector mode of Eq.~(\ref{eq:non-sep}). The half-wave plate HWP$_2$ is placed there for compensating slight changes that dielectric mirrors could introduce in the polarisation states. Note that all are unitary operations over the full system. Next, the vector mode from the entanglement stage, graphically shown in Fig~ \ref{fig:opticalexp}, is then passed to the \textbf{Work Extraction Stage.} At this point, Maxwell's demon has managed to sort the OAM states, based on their sign, into orthogonal polarisation states, a fact that can be used to extract useful work from the system. Therefore, we implement another conditional operation using now the PBS$_2$:  If the polarisation state is $\ket{H}$ keep the sign of the OAM but if the polarisation state is $\ket{V}$ flip the sign of the OAM. Applying this rule we create the separable state given by Eq. (\ref{eq:final}) which has the capacity to perform useful work. Additionally, we validated the OAM composition of the states at the end of each stage through OAM Measurement, which  was performed using a diffractive mode sorter composed of a phase unwrapper  (UW) which performs a conformal log-polar transformation, a phase corrector (PC) which corrects for inhomogeneous optical path length of the transformed beam, and a Fourier transforming lens (FTL) for sorting the OAM into lateral spots in the far field depending on the inserted OAM superpostion mode \cite{filippo2018}, as shown in the purple box labeled as ``OAM Measurement'' in Fig.~\ref{fig:opticalexp}, a more detailed explanation about the working principle of the mode sorter can be found in the supplementary. Lastly, polarisation measurement at the end of each stage was performed using Stokes polarimetry, which was spatially resolved through a polarisation camera and the QWP$_2$, as depicted in the ``Polarisation Measurement'' box in the same figure.

\section{Results}\label{sec:results}

\begin{figure*}[t]
\centering
\includegraphics[width=1\linewidth]{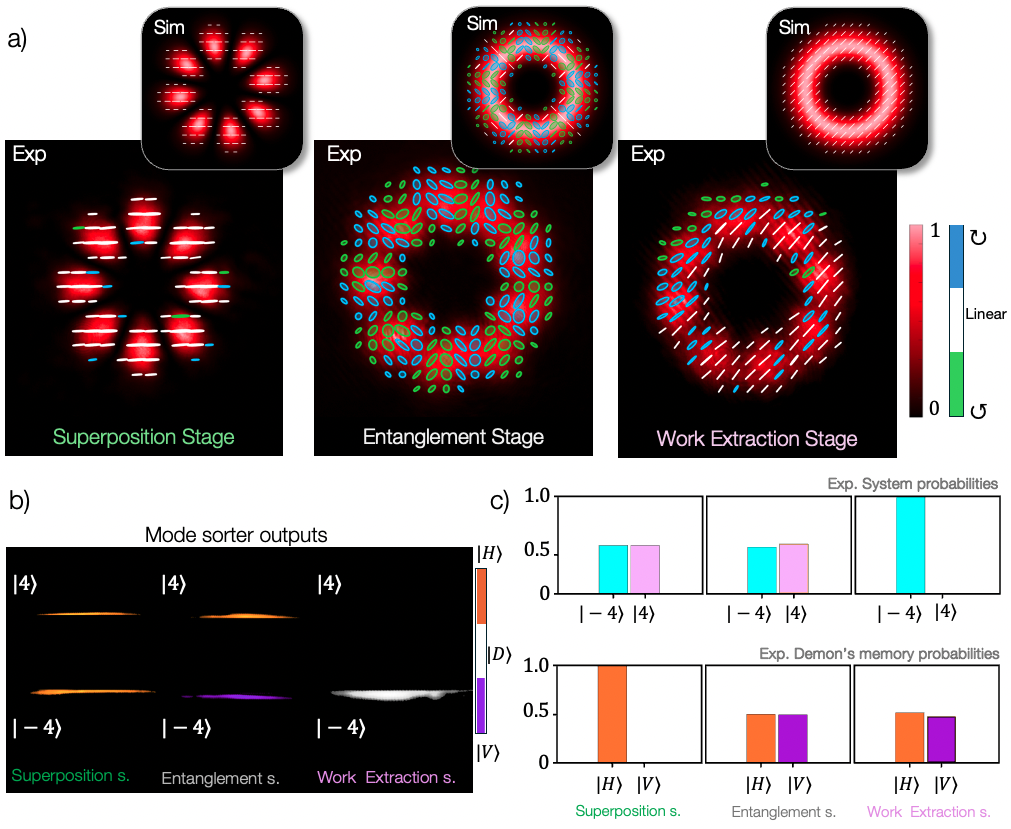}
\caption{ \textbf{Experimental results of the angular momentum  Maxwell's demon.} \textbf{a}, Experimental results of each state. The far-left snapshot shows an experimental state with two indistinguishable orbital angular momentum (OAM) modes, $\ket{\ell=4}$ and $\ket{\ell=-4}$, encoded with the same polarisation $\ket{H}$ (represented by horizontal lines). In the center, the entanglement state is depicted. At the far-right is the work extraction state. At each stage, the corresponding theoretical simulations are shown at the top. \textbf{b}, mode sorter outputs at each stage. \textbf{c}, Experimental probabilities for system and demon's memory.}
\label{fig:results}
	\end{figure*}

An essential characteristic of our Maxwell's demon is  its ability to extract work from a system by first entangling  itself with the system memory, thereby increasing the entropy locally and thereafter reducing it. Our experiment gives full access to the density matrix $\rho_d$ of the demon's memory, including its von Neumann entropy $S_d= \Tr(\rho_d \ln \rho_d)$. We perform a complete reconstruction of $\rho_d$ using the measured Stokes parameters from which all other parameters can be inferred (see Supplementary Information for full details of what follows).

At the \textit{Superposition stage}, the system and demon begin in a product state Eq.~(\ref{eqn:in}), with a concurrence value $C=0.1571 \pm 0.1304$. The system is in a scalar superposition of OAM, specifically $\frac{1}{\sqrt{2}}(\ket{-4} +\ket{4} )_s$ with $\theta=0$. Meanwhile, the demon’s memory state is characterized by a determinate state of horizontal polarization $\ket{H}$, as shown in Fig.~\ref{fig:results}(a). 

%the system begins in a superposition, where maximum entropy is expected, $S_s = 1$, calculated from the indistinguishable beams with $\ket{\ell=4}$ and $\ket{\ell=-4}$. Meanwhile, the demon state starts in a determined state of polarization $\ket{H}$, with an experimental entropy $S_d=0.08$. 
Subsequently, at the \textit{Entanglement stage} the action of the demon lead us to a entangled state (output state of Eq.~[\ref{eq:non-sep}]), the experimental concurrence obtained in this case is $C=0.9985\pm 0.0004$. Both the system and the demon's memory are in superposition states as shown in the central image of Fig.~\ref{fig:results}(a). Finally, at the \textit{Work extraction stage}, work extraction from the system is achieved by the conditional flipping of one OAM state, i.e., $\ket{\ell=4}\rightarrow \ket{\ell=-4}$ , according with the mapping in Eq~(\ref{eq:final}). As a result, the system undergoes two key outcomes:
\textit{i}) The energy cost associated with the OAM flip can be interpreted as work extraction, yielding $W\propto\vert \ell\hbar\vert= 4\hbar$ per photon. This quantifiable amount of extracted work corresponds to the angular momentum shift between the initial and final states, directly proportional to $\hbar$. \textit{ii}) The final state of this process results in a product state between the system and the demon, the experimental concurrence value is $C=0.2885\pm 0.2209$. This disentanglement marks the completion of the work extraction procedure, effectively decoupling the system from the demon's influence. In addition, we calculate the quantum information associated with each stage using von Neumann entropy. The entropies of both the demon and the system are computed using their respective reduced density matrices (refer to the Supplementary Information for a complete description).

Table \ref{table:comparison} summarizes the quantum information analysis of the process described above. It can be noted that the entropy of the entire system remains conserved throughout since only unitary transformations are applied. At the superposition stage, system and the demon's memory are in a separable state, meaning no quantum information is exchanged between them. In contrast, during the entangled stage, both the system and the demon's memory experience an increase in their respective entropies. Our results shows the critical role of entanglement as a resource that the demon can use not only to associate a polarisation state with each OAM but also to allow the demon to perform subsequent work extraction.

%%%
\begin{table}[]
    \centering
    \begin{tabular}{ |p{2.4cm}|p{2.4cm}|p{2.6cm}| }
 \hline
 \multicolumn{3}{|c|}{Experimental results quantum entropy ($S$)} \\
 \hline
Stage   & Demon's quantum information &  System's entropy/ information \\
 \hline
  \hline
Superposition  &  $0.0731 \pm 0.0097$ &  $0.00 \pm 00$  \\
 \hline
 Entanglement &$0.9999 \pm 0.0035$ & $0.9985 \pm 0.0055$ \\
 \hline
Work Extraction & $0.00 \pm 0.00$  & $0.0433 \pm 0.0096$ \\
 \hline
\end{tabular}
    \caption{The von Neumann entropy for the superposition, entanglement, and work extraction stages of the demon's memory and the system is examined. Initially, the demon's memory has an entropy of $S_d=0.073$, representing a determined state. At the entanglement stage, the demon's memory increase its entropy  $S_d=0.999$. At the last stage, the demon's memory results in a superposition state, with its quantum information being  zero. The system's entropy, follows the same behavior in terms of quantum information.}
   \label{table:comparison}
\end{table}

%%%%%%

%\begin{table}[ht]
%\begin{tabularx}{0.48\textwidth} { 
%   >{\raggedright\arraybackslash}X 
%   >{\centering\arraybackslash}X 
%    >{\centering\arraybackslash}X
%   >{\raggedleft\arraybackslash}X  }
% \hline\hline
%       Stage   &Full System  quantum information    & Demon's quantum information &  System's entropy/ information \\ [0.7ex] 
% \hline
%Superposition & $0.0959 \pm 0.0207$ & $0.073$ &  \IN{$0.056$}  \\ 
 
%Entanglement & $0.9999 \pm 0.0004$ &$0.999$ & $0.998$ \\

%Work extraction & $0.2497 \pm 0.0655$ &\IN{ $0.068$}  & $0.043$ \\
%\hline
%\end{tabularx}
% \label{table:comparison}
% \caption{The von Neumann entropy for the superposition, entanglement, and work extraction stages of the demon's memory and the system is examined. Initially, the demon's memory has an entropy of $S_d=0.073$, representing a determined state. At the entanglement stage, the demon's memory increase its entropy  $S_d=0.999$. At the last stage, the demon's memory results in a superposition state, with its quantum information being approximately zero. The full system and the system's entropy, follows the same behavior in terms of quantum information.}
%\end{table}

In addition, we conducted a classical entropic ($\Tilde{S}$) analysis of the entire process by observing the probability distributions throughout the three stages,  for both system and demon's memory. The experimental probabilities were  derived from the Stokes parameters for the demon's memory, as well as from the spatially distributed intensity in the case of the OAM system (see Supplementary Information). The probability distributions are shown in Fig.~\ref{fig:results}(c).  Initially, at the \textit{Superposition stage}, the system starts with equally distributed probabilities associated to the superposition state of OAMs, being the probabilities $P(\ket{-4}_s)=0.5260 \pm 0.0057$ and $=P(\ket{4}_s)=0.4740 \pm 0.0040$, the entropy of this classical probability distribution is  $\Tilde{S}_s=0.9911$ reflecting the superposition nature of the state. The demon's memory, starts with probabilities $P(\ket{H}_d)= 0.9911 \pm 0.0097$ and $P(\ket{V}_d)= 0.0089 \pm 0.0097$, so as a well defined state  its entropy is equal to $\Tilde{S}_d=0.073$. At the end of the process, we can see from  Fig.~\ref{fig:results}(c), that the system is in a well determinate state with $P(\ket{-4}_s)= 0.0047 \pm 0.0001$ and $P(\ket{4}_s)=0.9953 \pm 0.0098$. The entropy in this case reflects the full knowledge of the state of the system with $\Tilde{S}_s=0.0431$. In contrast, due the sorting and work extraction process the demon increased its entropy, with a value $\Tilde{S}_d=0.999$, calculated from the probabilities at the final stage $P(\ket{H}_d)= 0.5185 \pm 0.0149$ and $P(\ket{V}_d)= 0.4815 \pm 0.0149$. Table \ref{table:classentropy} summarizes the classical entropy at each stage, illustrating the transference of superposition from the system to the demon memory  as a result of the sorting processes and subsequent work extraction. The intermediate probabilities are included for completeness, both the system and the demon's memory are in indistinguishable states, reaching maximum of entropy.
 
 \begin{table}[]
    \centering
    \begin{tabular}{ |p{2.4cm}|p{2.4cm}|p{2.6cm} |}
 \hline
 \multicolumn{3}{|c|}{Experimental results classical entropy $(\Tilde{S}$)} \\
 \hline
 Stage & System  & Demon's memory \\
 \hline
  \hline
Superposition  & $0.9911\pm 0.018$    & $ 0.073\pm 0.0097$  \\
 \hline
 Entanglement & $0.9999 \pm 0.0004$ & $0.9999 \pm 0.0035$ \\
 \hline
Work Extraction &  $0.0433 \pm 0.0096$  & $0.999 \pm 0.000$ \\
 \hline
\end{tabular}
    \caption{Classical information is obtained at each stage using Shannon's entropy $(\Tilde{S}$). The indistinguishability of the system at the initial stage is transferred to the demon's memory at the end of the process.  }
    \label{table:classentropy}
\end{table}

\section{Discussion and Conclusion}

The concept of Maxwell's demon revolutionized classical thermodynamics, by proposing a link between information and thermodynamic work. This idea not only questioned the inviolability of the second law but also suggested that information could be harnessed to convert energy, reshaping our understanding of these fundamental concepts. By acquiring information about a system through measurements and precise control, the system can be manipulated to extract its energy. This concept also applies to its quantum version, with some key differences. Unlike the classical case, which involves a large number of particles, a quantum analogy can be made using a simple two-level system for both the system and the demon's memory \cite{lloyd97}. Additionally, quantum resources, like entanglement, enable the measurement to not only extracts the system’s information but also transform its state in the process.

Thus, we have proposed a simple protocol to simulate a quantum Maxwell's demon by exploiting the angular momentum degrees of freedom of non-quantum light. It is shown that non-separability is the key resource in the sorting process and work extraction. The energy extracted in this process by the flipping one of the OAMs states of the system can be seen as an information driven optical spanner, facilitating the rotation of physical objects by information. Our demon can then be considered as a simple optical machine, measuring orbital angular momentum by using polarisation. Initially, a product state is formed by system and demon's memory. Upon applying a conditional operation, the initial state becomes non-separable. Our system relies on the non-separability of angular momentum degrees of freedom to perform the sorting process. Having established the sorting stage, we introduced a strategy to extract work from the system by selectively flipping one of its the OAM states. Although we used a protocol where the system was encoded in the OAM degrees of freedom and the demon had a two-state SAM  memory as an example of our optical simulator, this proposed protocol can also be realized encoding the system state in the SAM and the demon in OAM.

Summarising, by starting with a system in a superposition state, we were able to harness the demon to analyze the classical information or entropy evolution of the system and the demon's memory, showing that a decrease in the system's entropy over the process is compensated by the increase in the demon's memory, meanwhile in terms of the quantum information we can see a conservation of the entropy in the overall process, ensuring no violation of second law occurs. Therefore, we believe that an angular momentum-based Maxwell's demon can potentially pave the way for new experimental light-control processes in quantum information research.

\section*{Acknowledgements}
The authors would like to acknowledge the support of South African Quantum Technology Initiative (SA QuTI), Department of Science and Innovation (South Africa), and the National Research Foundation (South Africa).
Edgar Medina-Segura (CVU 742790) would also like to acknowledge funding from CIO and CONAHCYT.
%\appendix
%\section{Supplementary information}\label{section:SI}

\section{Supplementary Information}
\subsection{Entropy analysis }\label{section:SI}

\subsubsection{Von Neumann's entropy}\label{subsection:VNE}

Von Neumann's entropy of a quantum state $\rho$ can be expressed as
\begin{align}\label{eq:entropy}
    S(\rho)\equiv -\Tr(\rho \log_2 \rho)=-\sum_x \lambda_x \log \lambda_x ,
\end{align}
being $\lambda_x$ the eigenvalues of $\rho$. It is clear from the definition that the entropy is zero if and only if the state is pure, and  reaches its maximum $\log d$ for a completely mixed state $I/d$. For a detailed explanation of entropy and its properties, the reader can refer to \cite{nielsen_chuang_2010}.

In our experiment, we can calculate the density matrix of each state and their respective  entropies. A summary can be found in Table \ref{table:entropy}. The density matrices at each stage are given by\\

\textit{a) Superposition stage}. The full system has a density matrix given by 
    \begin{align}
     \rho_{sd}=\rho_{s}\otimes \rho_{d}
     \end{align}
 For separable states, the total von Neumann entropy is equal to the sum of the entropies of the individual subsystems $S( \rho_{sd})=S( \rho_{s})+S( \rho_{d}) =0$. Thus, the reduced density matrices of the subsystems are
 \begin{itemize}
     \item System density matrix 
     \small{
     \begin{align}\label{eq:oam}
     \rho_s=\frac{1}{2}( \ket{\ell}\bra{\ell}+\ket{\ell}\bra{-\ell}+\ket{-\ell}\bra{\ell}+\ket{-\ell}\bra{-\ell})
     \end{align}}
     with eigenvalues $\lambda_1=1$ and $\lambda_2=0$, $S(\rho_s)=0$. 
     \item Demon's memory density matrix
     \begin{equation}
          \rho_d=\ket{H}\bra{H}
     \end{equation}
     is in a determinate state , with $\lambda_1=1$ and $S(\rho_d)=0$ .
 \end{itemize}
    
\textit{ b) Entanglement stage}. The density matrix of the composed system
    \begin{align*}
         \rho_{sd}&=\frac{1}{2}(\ket{\ell}\ket{H}\bra{\ell}\bra{H}+ \ket{-\ell}\ket{V}\bra{\ell}\bra{H}\\ \nonumber
         &+\ket{\ell}\ket{H}\bra{-\ell}\bra{V}+ \ket{-\ell}\ket{V}\bra{-\ell}\bra{V})
    \end{align*}
%On the other hand the demon's memory is in a determinated state with an unique $\lambda=P(\ket{H})=1$, thus  $\rho_d=\lambda\ket{H}\bra{H}$.    
By the partial trace, we calculate the reduced density matrices of the system and demon 
\begin{itemize}
    \item System
    \begin{align*}
    \rho_s=\Tr_d(  \rho_{sd})=\frac{1}{2}(\ket{\ell}\bra{\ell} + \ket{-\ell}\bra{-\ell}) = I/2
\end{align*}
\item Demon's memory 
    \begin{align*}
    \rho_d=\Tr_s(  \rho_{sd})=\frac{1}{2}(\ket{H}\bra{H} + \ket{V}\bra{V})= I/2.
\end{align*}
\end{itemize}
In both cases, we obtain a maximally mixed state with a maximum of entropy achieved $S(\rho_d)=S(\rho_s)=1$. Since,  both system and demon have eigenvalues $\{\frac{1}{2}, \frac{1}{2}\}$ in their respective subspaces. In addition, $S(\rho_{sd})=1$, with eigenvalues $\{\frac{1}{2}, \frac{1}{2}\}$ in the composed space. \\

\textit{c) Work extraction stage}. The density matrix for the full system is 
\begin{align}
     \rho_{sd}=\rho_{s}\otimes \rho_{d}
     \end{align}
with entropy $S( \rho_{sd})=S( \rho_{s})+S( \rho_{d}) =0$, where 
 \begin{itemize}
 \item System density matrix
 \begin{align}
     \rho_s=\ket{\ell}\bra{\ell}
 \end{align}
 is in a determinate state , with $\lambda_1=1$ and $S(\rho_s)=0$.
 \item Demon's density matrix
 \begin{align}\label{eq:pol}
    \rho_d=\frac{1}{2}(\ket{H}\bra{H} + \ket{V}\bra{V}).
\end{align}
with eigenvalues $\lambda=\{1,0\}$ and $S(\rho_d)=0$.
 
 \end{itemize}

\begin{table}[]
    \centering
    \begin{tabular}{ |p{1.9cm}|p{1.9cm}|p{1.9cm}|p{1.9cm}|  }
 \hline
 \multicolumn{4}{|c|}{Theoretical results quantum entropy} \\
 \hline
 Stage & $S(\rho_s)$   & $S(\rho_d)$& $C(\rho_{sd})$\\
 \hline
  \hline
Superposition   & 0    &  0 &  0 \\
 \hline
 Entanglement & 1 & 1 &1\\
 \hline
Work Extraction &   0 & 0 &  0 \\
 \hline
\end{tabular}
    \caption{Theoretical results. Von Newmann's entropy $S_{\{ s,d\}}$ where the values of each stage are calculated using the eigenvalues $\lambda_i$  of the density matrix in each state. The concurrence of the full system  $\mathcal{C}(\rho_{sd})$ of each stage is displayed on the last column, note that initial and final stages are product states with zero entanglement while at the sorting state system and demon shows a maximum of non-separability.  }
    \label{table:entropy}
\end{table}

 Table \ref{table:entropy} summarizes the quantum entropy at each stage, illustrating the transference of superposition from the system to the demon memory  as a result of the sorting processes and subsequent work extraction. 
 
\subsubsection{Classical entropy}\label{ss:classical}

In addition, we can extract the classical information at each stage, using Shannon entropy to measure of uncertainty or information content in the probability distribution,  defined as
\begin{equation}
    \Tilde{S}= -\sum_{j}p_j\log_2(p_j)
\end{equation}
being $p_j$ the probability associated to the state $\ket {\psi}_j$. Higher entropy values indicate more uncertainty, similarly to the quantum information case.  In our case, the theoretical results can be described as follows 

\textit{a) Superposition stage}. The state vector of the system and demon are given by
 \begin{itemize}
     \item Superposition system state 
     \small{
     \begin{align}
     \ket{\phi_i}_s=\frac{1}{\sqrt{2}}( \ket{\ell}+ \ket{\ell})
     \end{align}}
  representing a superposition state  with probabilities $p_1=p_2=1/2$, and $\Tilde{S}(\ket{\phi_i}_s)=1$. 
     \item Determinate demon's memory state
     \begin{equation}
          \ket{\psi_i}_d=\ket{H}
     \end{equation}
    with $p_1=1$, and $\Tilde{S}(\ket{\psi_i}_d)=0$.
 \end{itemize}
    
\textit{ b) Entanglement stage}. System and demon's memory are in a superposition state, where their density matrices $\rho_s=\rho_d=I/2$, representing a  classical state (i.e., it can be described by a probability distribution over distinct states). von Neumann's entropy reduces to Shannon's entropy in this case \cite{nielsen_chuang_2010}, thus
\begin{align*}
S(I/2)=\Tilde{S}(I/2)=1
\end{align*}
\\

\textit{c) Work extraction stage}. The vector states of system and demon are 
 \begin{itemize}
 \item System state
 \begin{align}
     \ket{\phi_f}_s=\ket{\ell}
     \end{align}
 is in a determinate state, with $p_1=1$ and $\Tilde{S}(\ket{\phi_f}_s)=0$.
 \item Superposition demon's memory state
 \begin{align}\label{eq:pol}
    \ket{\psi_f}_d=\frac{1}{\sqrt{2}}(\ket{H}+ \ket{V}).
\end{align}
with probabilities  $p=\{1/2,1/2\}$ and $\tilde{S}( \ket{\psi_f}_d)=1$.
 
 \end{itemize} 
\begin{table}[]
    \centering
    \begin{tabular}{ |p{2.4cm}|p{2.4cm}|p{2.6cm} |}
 \hline
 \multicolumn{3}{|c|}{Theoretical results classical entropy} \\
 \hline
 Stage & System  & Demon's memory \\
 \hline
  \hline
Superposition  & 1    &  0  \\
 \hline
 Entanglement & 1  & 1 \\
 \hline
Work Extraction &  0  & 1  \\
 \hline
\end{tabular}
    \caption{Classical information is obtained at each stage using Shannon's entropy. The indistinguishability of the system at the initial stage is transferred to the demon's memory at the end of the process.  }
    \label{table:classentropy}
\end{table}

Table \ref{table:classentropy} summarizes the classical entropy at each stage, illustrating the transference of superposition from the system to the demon memory  as a result of the sorting processes and subsequent work extraction. 
\begin{figure*}[t]
\centering
\includegraphics[width=1\linewidth]{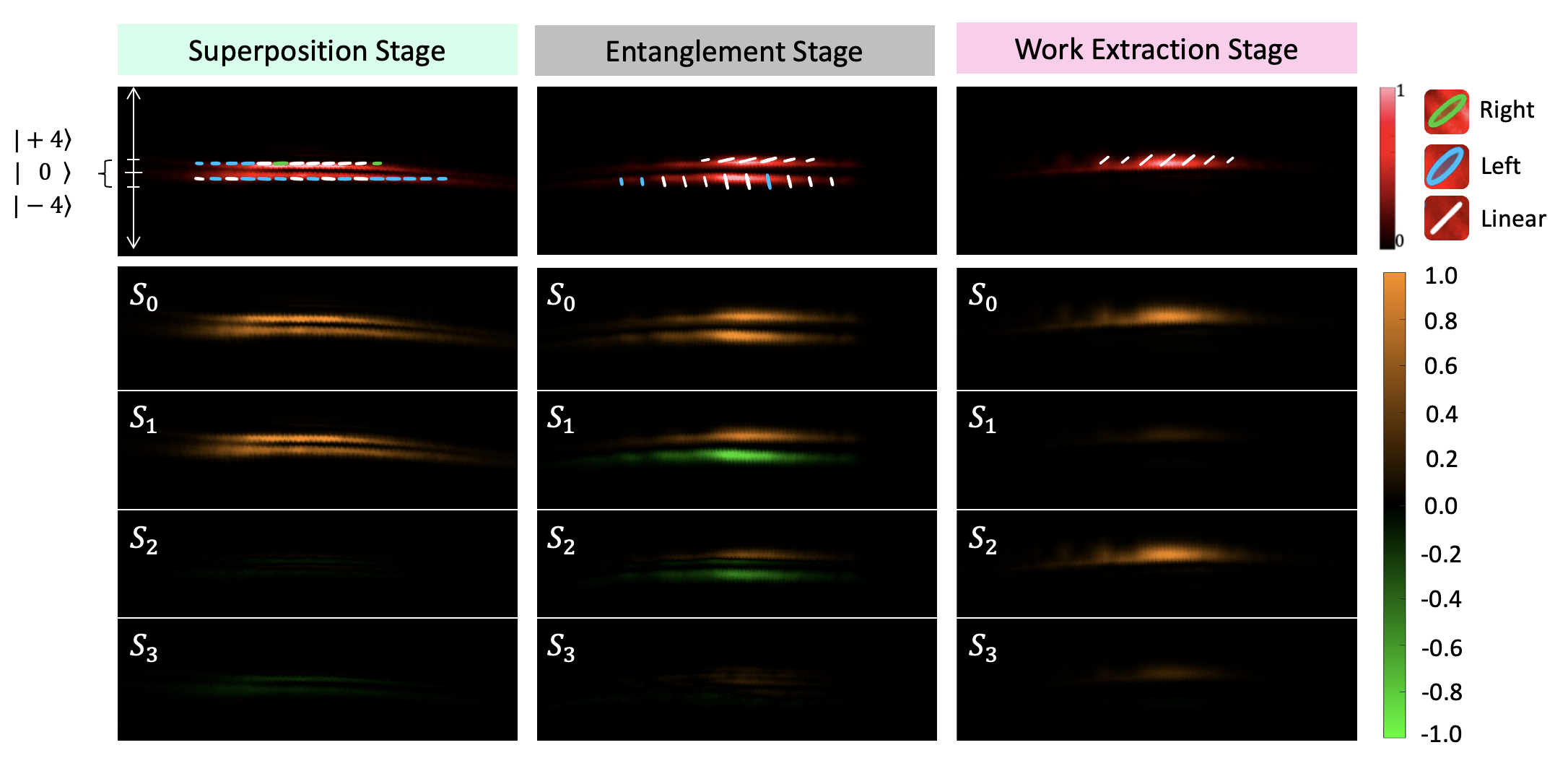}
\caption{Experimental Results. A demon's memory  and system 
 are  encoded in polarization $\{\ket{H},\ket{V}\}$ and OAM  $\{ \ket{-4},\ket{4}\}$, respectively. At the \textit{Superposition state}: The demon's memory is initialized in the $\ket{H}$polarization, represented by the horizontal lines in the top-right snapshot. The two OAMs are identified at the OAM detection stage according to the calibration described in Sec.\ref{section:Mode sorter}. At the bottom, the Stokes parameters reveal the detection of only horizontal polarization. At the \textit{Entanglement stage}: A non-separable state is formed between the system and the demon's memory. The respective Stokes parameters are shown at the bottom. At the \textit{work extraction stage}: The demon's memory ends in a superposition of $\ket{H}+\ket{V}$, depicted in the snapshot by the diagonal linear polarization. The Stokes parameters for this stage are also shown at the bottom.}
\label{fig:stokes}
	\end{figure*}
\subsection{Stokes parameters }\label{section:SI}
Stokes polarimetry spatially resolved allows the measurement of the polarisation states along the transversal plane of a mode. Stokes parameters can be calculated experimentally from six intensity measurements \cite{goldstein2011}
%\cite{Selyem2019,Singh20}:
\begin{align}
   \mathcal{S}_0 &= I_H + I_V,\\ \nonumber
   \mathcal{S}_1 &= I_H - I_V,\\ \nonumber
   \mathcal{S}_2 &= I_D - I_A,\\ \nonumber
   \mathcal{S}_3 &= I_R - I_L,
\end{align}

where $I_H$, $I_V$, $I_D$, $I_A$, $I_R$ and $I_L$ represent the intensity of the horizontal, vertical, diagonal, anti-diagonal, right-circularly and left-circularly polarization components, respectively. Figure \ref{fig:stokes} shows the stokes parameters for superposition, entanglement and work extraction stage in our experiment measured after the Mode sorter for which we can associate vertical position of these elongated spots with OAM. Once the Stokes parameters have been determinated, the reconstruction of the polarized state is determinated by \cite{nape2023quantum}.

\begin{align}
    \rho_{pol} =\frac{1}{2}(I + \sum_i \mathcal{S}_i\cdot\sigma_i) 
\end{align}

The reader may find the reconstruction of the state by polarisation familiar, as it resembles the structure of a qubit. This equivalence is given by ($\mathcal{S}_i \equiv r_i$), where the Stokes parameters represent the observables in the polarization subspace, analogous to how  $r_i$ corresponds to the Bloch vectors representing the observables of a qubit state \cite{nape2023quantum}.

 Stokes polarimetry can also be used to quantify the amount of non-separability of a classical optical field. To measure the vector nature (``or vectorness") of a beam, we use the measure of concurrence \cite{Ndagano16,Selyem2019,wootters2001} defined as

\begin{align}
 \mathcal{C} =\sqrt{1- \sum_{i=1}^{3}\frac{
 \mathcal{S}^{2}_{i}}{\mathcal{S}_0^2}}
\end{align}

A maximum entangled state is found when $\mathcal{C}=1$, while for product states or separable states we found $\mathcal{C}=0$. For the entropy calculations, the eigenvalues can be expressed in terms of the stokes vector $r=\sum_{i=1}^{3}\frac{\mathcal{S}^{2}_{i}}{\mathcal{S}_0^2}$ or the concurrence as follows:

\begin{equation}\label{eq:eigenvalues}
    \lambda_{1,2}=\frac{1\pm r}{2}=\frac{1\pm (1-C^2)}{2}
\end{equation}
and calculate the entropy using Eq. [\ref{eq:entropy}].
\subsection{Mode sorter}\label{section:Mode sorter}

\begin{figure*}[t]
\centering
\includegraphics[width=0.9\linewidth]{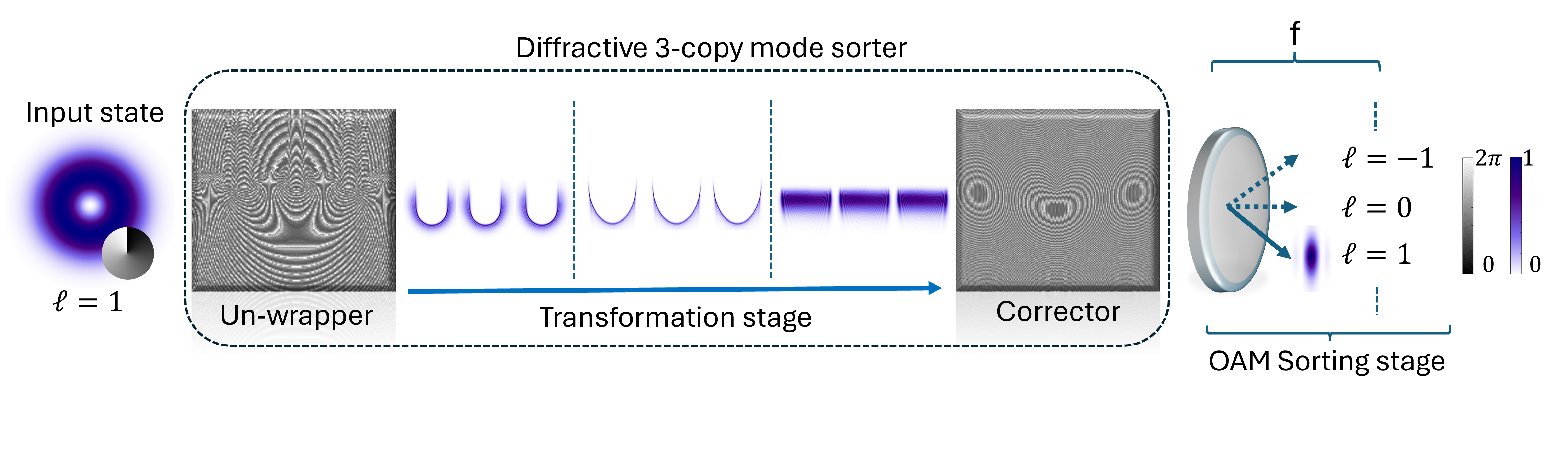}
\caption{\textbf{Concept of mode sorter}: This schematic displays a simulation of how the un-wrapper phase profile unravels a given OAM input mode into transverse position (three rectangular strips) through propagation to the corrector phase term, the corrector phase term then corrects the field to maintain the shape of the unravelled mode and a Fourier transforming lens is used to sort the field based on the OAM in the field, for example here, we used $\ell=1$ and thus the Fourier transformed field will land on the postion where OAM of $\ell=1$ is expected as shown above}
\label{fig:concept}
\end{figure*}

Orbital angular momentum (OAM) mode sorters are optical instruments that can be used to characterise  light beams with a phase cross section azimuthal term exp$(i\ell\theta)$ carrying orbital angular momentum (OAM) of $\ell \hbar$ per photon, where $\ell$ and $\theta$ are the azimuthal index and can take any integer value \cite{allen1992orbital,molina2007twisted,franke2008advances,forbes2024orbital} and the azimuthal angle respectively. These optics implement a conformal geometrical transformation that
enables spatial separation of OAM states and their superposition, this geometrical transformation in the context of mode sorters is the traditional log-polar transform \cite{berkhout2010efficient}, whereby a Laguerre-Gaussian (LG) beam with a helical phase front is transformed into a transverse position by applying a phase term called an un-wrapper $\phi_1(x,y)$ as described by 
\begin{small}
    \begin{equation}
    \phi_1(x, y) = \frac{2 \pi a}{\lambda f} \left[ y \arctan \left( \frac{y}{x} \right) - x \ln \left( \frac{\sqrt{x^2 + y^2}}{b} \right) + x \right],
    \label{eq:1}
\end{equation}
\end{small}

where $ \lambda $ represents the wavelength of the incoming beam, $f = 8.5 mm$ is the focal length of the Fourier transforming lens, $a = d/\pi$ ($a = 500/2\pi$), whereby $d$ is the length of the transformed beam, $b=900\mu$ translates
the transformed beam in the $u$ direction. This corresponds to the transformation of an
input image containing concentric circles into an output
image of parallel lines. Transforming each input circle onto an
output line provide the required deviation in ray direction and
therefore the phase profile of the transforming optical element. The transformation introduces a phase distortion due to the resulting variation in optical path length in the transformation, which needs to be corrected by a second element described by
\begin{equation}
    \phi_2 (u, v) = -\frac{2 \pi a b}{\lambda f} \exp \left( - \frac{u}{a} \right) \cos \left( \frac{v}{a} \right).
    \label{eq:2}
\end{equation}

\begin{figure*}[t]
\centering
\includegraphics[width=0.9\linewidth]{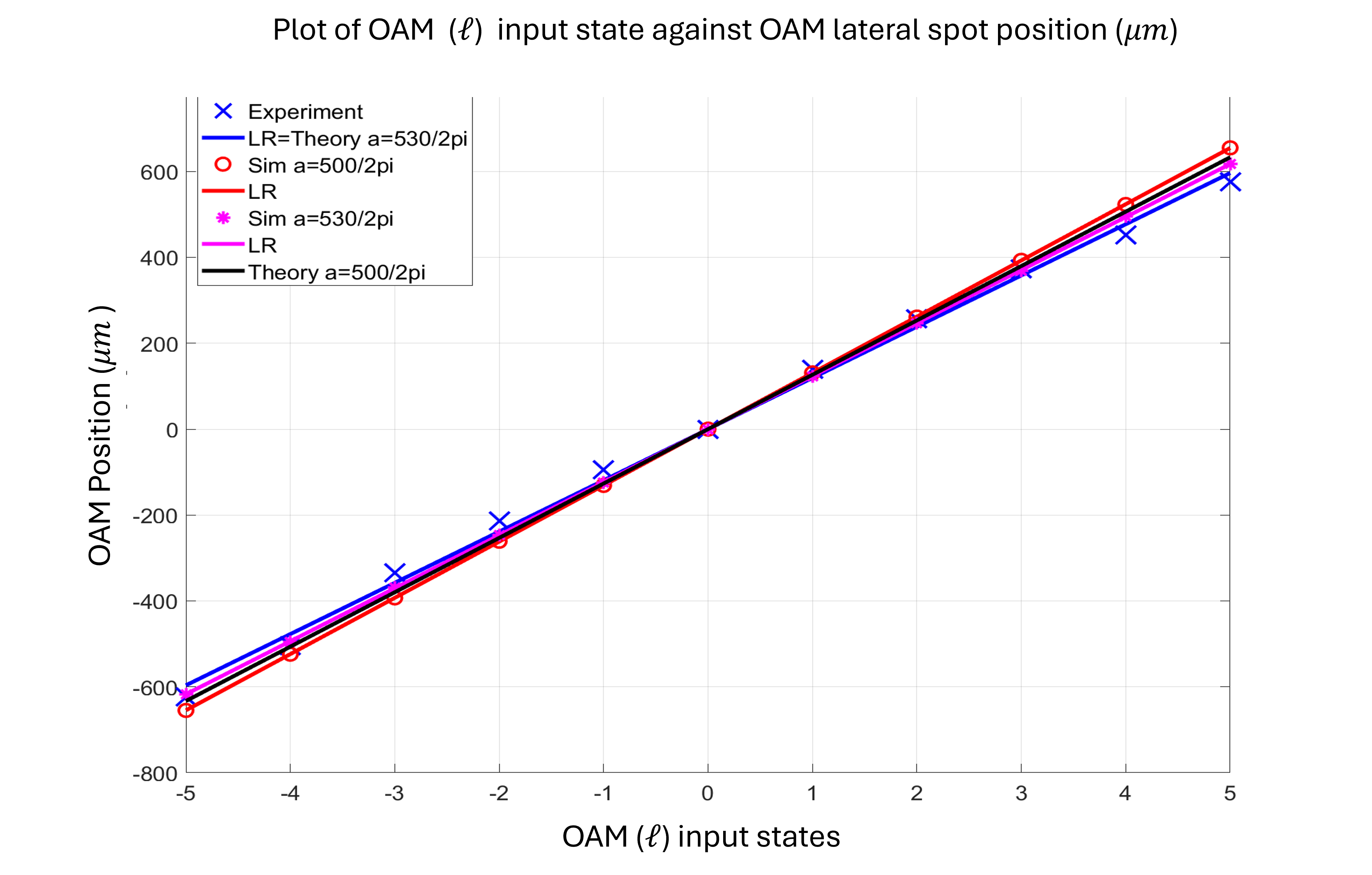}
\caption{Calibration plot of OAM lateral positions versus OAM input states comparing  the theoretical, simulated and experimental relations as per  $ t_{\ell} = \frac{\lambda f}{d} \ell$.}
\label{fig:oamvspos}
\end{figure*}

The transforming system therefore includes two custom optical
elements, one transforms the image using log-polar transform and a second, placed
in the Fourier plane of the first element, for the correction of phase distortion. The OAM states are then sorted by taking the
transformed transverse beam into the far-field via a lens. Different OAM beams (topological values $\ell$) are
sorted to different positions in the far-field, hence we characterize the system by plotting a position versus OAM (topological values, $\ell$) of the
theoretical, simulated and experimental relationship as seen in Fig.~\ref{fig:oamvspos}, governed by $ t_{\ell} = \frac{\lambda f}{d} \ell$. In our experiment we specifically used a 3-copy diffractive mode sorter which is advantageous to refractive mode sorter due to its lower cross-talk when sorting OAM modes. This directly means the un-wrapper phase term adds an extra term called a fan-out term which makes multiple copies (3 copies in our case) and a lens term, so the new diffractive un-wrapper phase term is given by:

\begin{equation}
    \tau(x, y) = \exp(i \phi_1)  \exp(i \Omega_{FO,N})  \exp(i \phi_{lens}).
    \label{eq: diff-un}
\end{equation}

Here $\exp(i \Omega_{F0,N}) = \sum_{m=-(N-1)/2}^{(N-1)/2} c_m e^{i (\gamma_m + \delta_m)}$ splits the unwrapped beam into $N$ with  $(N=3)$ copies and locates the several copies of the beam side by side on the
second optical element as visually represented by Fig.~\ref{fig:concept}. $c_m$, $\delta_m$ are optimisation constants for equal distribution of energy of the transformed beam and $\gamma_m =mLk/f$. $\phi_1$ is still the un-wrapper and $\exp(i \phi_{lens})$ is a lens phase term. 

% Requires: \usepackage{array}
\begin{table}[h]
    \centering
    \begin{tabular}{|c|c|c|c|c|c|}
        \hline
        $N$ & optimization constants & 0 & $\pm1$ \\
        \hline
        
        3 & $\delta_3$ & 0 & $-\pi/2$ \\
        \hline
         & $c_3$ & 1 & 1.329 \\
        \hline
       
        \hline
    \end{tabular}
    \caption{Table showing the optimisation parameters for the three copy mode sorter following the fan-out term $\exp(i \Omega_{F0,N})$}
    \label{tab:para}
\end{table}

We need a phase-correction term that will correct both the log-polar transformation and the fan-out process. An analytical formulation of the log-polar phase-corrector is available in the paraxial regime. However, selecting specific focal lengths and beam sizes requires a more precise computation of phase patterns beyond the Fresnel regime. The resultant diffracted field \( U \) from the angular spectrum diffraction theory \cite{li2007diffraction} can be represented as a convolution algorithm:
\begin{small}
    \begin{equation}
   \label{eq:U}
   U(u, v) = \mathcal{FT}^{-1}\left\{ \mathcal{FT}\left\{T(x, y)U_0(x, y)\right\} H_{AS}(f_x, f_y) \right\},
\end{equation}
\end{small}

where $\mathcal{FT}$ and $\mathcal{FT}^{-1}$ and are the Fourier transform and the inverse Fourier transform, respectively. And $H_{AS}$ is the
angular spectrum transfer function

\begin{equation}
    H_{\text{AS}}(f_x, f_y) = \exp\left[ikz\sqrt{1 - (\lambda f_x)^2 - (\lambda f_y)^2}\right].
    \label{placeholder_label}
\end{equation}

The new phase corrector term can be visually represented in Fig.~\ref{fig:concept} and mathematically expressed as follows
\begin{equation}
    \Omega_{\text{PC}}(u, v) = 2\pi - \arctan\left[U(u, v)\right].
    \label{eq: CORR}
\end{equation}

This can be numerically calculated for $U_0$ as an input Gaussian mode with a planar phase front and a beam waist
precisely chosen to shine the zone of interest of the first element. Finally, the phase-corrector is
given a tilt term to prevent the beam from overlapping with a possible zero-order contribution. To summarize, we start with a light field carrying orbital angular momentum (i.e LG modes) and apply the un-wrapper phase described by Eq.~(\ref{eq: diff-un}), whereby the field will undergo a log-polar transformation, the transformed field will have some distortion due to different path length during the propagation of the field and a phase corrector described by Eq.~(\ref{eq: CORR}) is used to correct the field distortions and finally a Fourier transforming lens is used to sort the different input state of OAM in the system.

\subsection{Error propagation}
Intensity images were measured with a Monochrome Polarisation Camera CS505MUP1 from Thorlabs at 12-bit depth which implies a direct uncertainty of $\Delta I_0=1/2^{13}$ for each pixel. However, we normalised the images by the maximum value of $S_0$ for which the uncertainty takes the value of $\Delta I=1/(2^{13}S_{0max})$ for each pixel. Next, error propagation for indirect measurements is computed following the general method explained in Ref. \cite{baird}. Let $f(x_1,x_2,x_3,...,x_k)$ be an indirect measurement function of $k$ direct measurements with $\Delta x_i$ the uncertainty of the $i-th$ direct measurement. Then, the uncertainty for the indirect measurement $f$ is given by
\begin{equation}
    \Delta f=\sum_{i=1}^{k}\Bigg|\frac{\partial f}{\partial x_i}\Bigg|\Delta x_i,
    \label{ec_errorPropagation}
\end{equation}
where $|\partial f/\partial x_i|$ is computed for the specific values used to obtain each measurement $f$. For instance,  we compute the uncertainty for Concurrence which is given by
\cite{concurrence_PRA_2019}
\begin{equation}
    C(\mathbb{S}_0,\mathbb{S}_1,\mathbb{S}_2,\mathbb{S}_3)=\sqrt{1-\frac{\mathbb{S}_1^2+\mathbb{S}_2^2+\mathbb{S}_3^2}{\mathbb{S}_0^2}},
\end{equation}
where
\begin{equation}
    \mathbb{S}_i=\sum_{x,y}S_i(x,y), \qquad i=0,1,2,3
\end{equation}
being $S_i(x,y)$ the Stokes parameters images. Then
\begin{equation}
    \Delta C=\sum_{i=0}^{3}\Bigg|\frac{\partial C}{\partial \mathbb{S}_i}\Bigg|\Delta \mathbb{S}_i,
\end{equation}
where, according to Eq.~(\ref{ec_errorPropagation}):
\begin{equation}
    \Delta \mathbb{S}_i=\sum_{x,y}\Delta S_i(x,y), \qquad i=0,1,2,3,
\end{equation}
with the uncertainty for each pixel of the Stokes parameters images given by:
\begin{equation}
    \Delta S_i(x,y)=2\Delta I=\frac{1}{2^{12} S_{0max}},
\end{equation}
and
\begin{equation}
    \frac{\partial C}{\partial \mathbb{S}_0}=\frac{\mathbb{S}_1^2+\mathbb{S}_2^2+\mathbb{S}_3^2}{\mathbb{S}_0^3\sqrt{1-\frac{\mathbb{S}_1^2+\mathbb{S}_2^2+\mathbb{S}_3^2}{\mathbb{S}_0^2}}},
\end{equation}

\begin{equation}
\frac{\partial C}{\partial \mathbb{S}_j}=-\frac{\mathbb{S}_j}{\mathbb{S}_0^2\sqrt{1-\frac{\mathbb{S}_1^2+\mathbb{S}_2^2+\mathbb{S}_3^2}{\mathbb{S}_0^2}}}, \qquad j=1,2,3.
\end{equation}

Therefore, uncertainty for Concurrence is given by:
\begin{equation}
    \Delta C=\frac{|(1-C^2)\mathbb{S}_0|\Delta \mathbb{S}_0+|\mathbb{S}_1|\Delta \mathbb{S}_1+|\mathbb{S}_2|\Delta \mathbb{S}_2+|\mathbb{S}_3|\Delta \mathbb{S}_3}{\mathbb{S}_0^2C}.
\end{equation}

As another example, for the Von Neumann’s entropy Eq.~(\ref{eq:entropy}), with uncertainty given by 
\begin{equation}
    \Delta S(\rho)=\Bigg|\frac{\partial S}{\partial \lambda_1}\Bigg|\Delta \lambda_1+\Bigg|\frac{\partial S}{\partial \lambda_2}\Bigg|\Delta \lambda_2.
\end{equation}
The eigenvalues $\lambda_i$, Eq.~(\ref{eq:eigenvalues}) have the uncertainties 

\begin{equation}
    \Delta \lambda_{1,2}=\Bigg|\frac{\partial \lambda_{1,2}}{\partial C}\Bigg|\Delta C=C \Delta C.
\end{equation}
Consequently, 

\begin{align} \nonumber
    \Delta S(\rho)=&\frac{1}{\ln2}(|-\ln(\lambda_1)+1|\Delta \lambda_1+ |-\ln(\lambda_2)+1|\Delta \lambda_2).
\end{align}

The calculation of uncertainties for the rest of indirect measurements is analogous.

\end{document}